\documentclass[12pt]{iopart}

\usepackage{amsgen}
\usepackage{amssymb}
\usepackage{setstack}
\usepackage{bbold}
\usepackage{mathrsfs}
\usepackage{graphicx}
\usepackage{multicol}
\usepackage{color}

\def\be{\begin{equation}}
\def\ee{\end{equation}}
\def\bea{\begin{eqnarray}}
\def\eea{\end{eqnarray}}

\begin{document}

\title{
Black Holes as Echoes of Previous Cosmic Cycles}


\author{Bernard Carr$^1$, Timothy Clifton$^1$ and Alan Coley$^{2}$ }
\address{$^1$ Astronomy Unit, Queen Mary University of London, London E1 4NS, UK.\\
$^2$ Department of Mathematics  and Statistics, Dalhousie University, Halifax, Nova Scotia B3H 3J5, Canada.}
\eads{\mailto{b.j.carr@qmul.ac.uk, t.clifton@qmul.ac.uk, aac@mathstat.dal.ca}}

\begin{abstract}
The existence of exact solutions which represent a lattice of black holes at a scalar-field-dominated cosmological bounce suggests that black holes could persist through successive eras of a cyclic cosmology. Here we explore some remarkable cosmological consequences of this proposal. In different mass ranges pre-big-bang black holes could explain the dark matter, provide seeds for galaxies, generate entropy and even drive the bounce itself. The cycles end naturally when the filling factor of the black holes reaches unity and this could entail a dimensional transition. \\

\noindent {\it Essay written for the Gravity Research Foundation 2017 Awards for Essays on Gravitation. 
Submitted 31 March 2017. }
 \end{abstract}

In many cosmological scenarios, the current expansion phase of the universe would have been preceded by a collapse phase. The  bounce which connects these two phases necessarily violates the null energy condition and could derive from: (i) classical effects associated with a cosmological constant \cite{lemaitre}, a scalar field \cite{star}, higher-derivative effects \cite{novello} or other modified theories of gravity \cite{tors1}; (ii) semi-quantum gravitational effects associated with string theory \cite{string}, loop quantum gravity \cite{sing} or the pre-big-bang scenario \cite{BHScorr}; (iii) quantum gravitational effects associated with quantum gravity condensates \cite{oriti} or minisuperspace \cite{gielen}.
For a topical review of bouncing cosmologies, see Ref.~\cite{brand}. There have also been claims that scale-invariant fluctuations generated during a cosmological collapse phase could turn into scale-invariant curvature fluctuations after a bounce \cite{cai}. 
\if
[Recently Ijjas and Steinhardt  have considered whether a classical bounce can occur in models with a scalar field 
 with a cubic Galileon action \cite{is}. 
although this does not work with a quadratic potential. A ghost field (with negative kinetic term) is quantum mechanically unstable. 
On a homogeneous FRW background with $\rho >0$, any leading order instability comes from the kinetic ir gradient terms of the the linesr theory. 
They use a 2nd order action for comoving curvature perturbation and the inverse method to show that a classically stable non-singular bounce is possible.]
\if
The sound speed remains continuous and positive and below $1$, as required. 
In Refs.~\cite{deffayet,easson}, it was argued that cubic Galileons can smoothly pass through the bounce but encounter gradient instabilities before exiting the NEC-violating stage. Ref.~\cite{elder} showed that the scalar peturbations are stable during the early stages when $H$ is small and gravity negligible 
but did not include the the exit from the NEC-vioating phase when 
$H$ is large and the scalar field creates curvature perturbations with 
there are ghost and gradient instabilities. Such instabilites were explicitly encountered in Refs.~\cite{qiu,koehn}. Refs.~\cite{lib,kob} claims that singularities are unavoidable in NEC-violating Galileon theories. ]
\fi
If the bounce occurs due to quantum cosmological effects, then it presumably occurs at the Planck density.
The possibility of bouncing universes in the context of the no boundary wave function (NBWF) of quantum cosmology \cite{hh} has been considered by Hartle and Hertog \cite{hertog}.
Arrows of time are viewed four-dimensionally as properties of the four-dimensional Lorentzian histories of the universe and probabilities for these histories are predicted by the NBWF. 
For recollapsing classical histories with big bang and big crunch singularities the NBWF approach predicts that the fluctuations are small near one singularity and grow through the expansion and recontraction to the other singularity. The arrow of time defined by the growth in fluctuations 
is thus bidirectional in a bouncing spacetime.]
\fi

As stressed in a previous paper \cite{cc}, henceforth C$^2$, it is interesting to consider whether black holes can persist through a cosmological bounce and what effect they would have on the large-scale dynamics and on each other.
One {can} divide {such} black holes into two classes: (i) ``pre-crunch black holes'' (PCBHs) that persist in a universe that recollapses to a big crunch and then bounces into a new expansion phase; and (ii) ``big-crunch black holes'' (BCBHs) that are generated by the high density of matter at the bounce itself. We use the term ``pre-big-bang black hole'' (PBBBH) to cover both these  possibilities. Those of class (i) could be very massive, like the ones which form as a result of stellar collapse or reside in galactic nuclei in the present universe. Those of class (ii) could be very small and emit quantum radiation. However, they would generally complete their evaporation {\it after} the bounce and so would probably be observationally indistinguishable from  the ``primordial black holes'' (PBHs) which formed after the big bang. 
Quintin and Brandenberger \cite{quintin} have recently studied how BCBHs can form from density fluctuations in a contracting universe. \if
They solve the cosmological perturbation equationsfor a general hydrodynamical fluid, describing the evolution of the  perturbations as a function of the fluid's equation of state forboth quantum and thermal initial fluctuations. 
They  derive a general condition for black hole collapse on sub-Hubble scales and follow Carr \cite{c1975} in using the Press-Schechter formalism to describe the black hole formation probability. 
For a fluid with a small sound-speed (dust), both types of
fluctuations grow in a contracting universe and the largest inhomogeneities that collapse
into BCBHs have the Hubble size and collapse well before reaching the Planck density.
For a radiation-dominated fluid, no black hole can form before the Planck
density. Thus only bouncing models in which a
radiation-dominated era begins early in the cosmological evolution avoid BCBH  formation.
\fi

In a recent paper \cite{ccc}, henceforth C$^3$, we derived some exact solutions  which describe a regular lattice of black holes in a cosmological background dominated by a scalar field at the bounce. 
More precisely, we presented an exact solution of the Einstein constraint equations, which can
be used to develop a 4-dimensional dynamical solution  
in which multiple distinct black holes propagate through the bounce. 
Scalar-field-driven bounces are the simplest  ones to consider 
 in a fully relativistic way and our results
 illustrate that there exist exact solutions in which multiple black holes persist through a bounce. 
The bounce can occur well below Planck density, so a classical approach is still legitimate. 
\if
In this paper, we {study} this scenario {in detail} by considering models that contain a regular lattice of black holes and a scalar field at the maximum compression. These restrictions are primarily motivated by mathematical simplicity, as we find that they {yield} a number of exact solutions to the Einstein constraint equations. The {associated} initial data {then prescribe} a unique evolution in both the expansion and collapse {directions}. They therefore constitute a {proper} cosmological model. Although in general the evolution will need to be determined numerically, the initial data itself can be used to analyse a number of the problems outlined above. In particular, the positions of the apparent horizons can be determined and used to calculate the distance between neighbouring black holes. This information can be used to determine the fraction of space that is filled by black holes as a function of the maximum energy density at the bounce. It also unambiguously demonstrates that black holes can persist through a cosmological bounce.

First we considered hyperspherical cosmological models.  We found that the model parameter $\kappa$ must lie within a given range of values if the black holes are to have positive mass and the magnitude of the scalar field is to be large enough to ensure a bounce.  A total of six configurations for multiple, regularly-arranged black holes were considered, and it was found that the upper and lower bounds on $\kappa$ generally increase with the number of black holes.  Black holes in a spatially flat cosmology and in the presence of a scalar field were also considered, and were found to allow time-symmetric solutions with positive mass.  We again found {an} upper bound above which black holes merge, but unlike the hyperspherical cosmologies we found no lower bound.
\fi
One feature of these solutions is that the number of black holes ($N$) is finite and relatively small. In a closed model, $N$ has the possible values $5, 8, 16, 24, 120, 640$.  The associated cell size in the current universe is therefore of order $N^{-1/3}$ times the particle horizon size, which necessarily exceeds $600$~Mpc. This does not resemble the situation in the  real universe, where the number of supermassive black holes (SMBHs) is of order $10^{10}$ (one per galaxy) and the number of intermediate mass black holes (IMBHs)  could be of order $10^{20}$ (if they provide the dark matter). However, $N$ could be arbitrarily large in a flat or open universe, so this may be more applicable. 

The emphasis of C$^3$ was mathematical, so the purpose of this essay is to examine some interesting cosmological applications of this result. 

\begin{itemize}

\item {\it PBBBHs as dark matter.}
The suggestion  that the dark matter could comprise PBHs has recently become popular. However, PBBBHs would be equally plausible and would have almost indistinguishable consequences. In this case, the same black holes would provide the dark matter in successive cosmic cycles, with the dark matter fraction progressively increasing. 
As discussed in Ref.~\cite{cks}, there are a wide variety of lensing, dynamical and astrophysical constraints 
for non-evaporating black holes (i.e. those  larger than $M_* \sim 10^{15}$g). 
There are four mass windows in which this is possible: around the Planck mass ($10^{-5}$g) if evaporating black holes leave stable relics rather than disappearing completely; the atomic-sized range ($10^{16} - 10^{17}$g); the sublunar mass range ($10^{20} - 10^{24}$g); and the intermediate mass range ($10 -10^3 M_{\odot}$). In the last case, it has been suggested that the binary black holes detected by LIGO could be primordial but they could also be PBBBHs.  \\

\item {\it PBBBHs as seeds for galaxies.}
It is known that most galactic nuclei  contain supermassive black holes (SMBHs), extending from $10^{6}M_\odot$ to $10^{10}M_\odot$ and already in place by a redshift $\sim 10$. However, it is hard to understand how such enormous black holes could have formed so early unless there were already large seed black holes well before galaxy formation
\cite{dolgov}. Indeed, pregalactic SMBHs could act as condensation nuclei for galaxies through their gravitational Coulomb effect \cite{carr-rees}. 
The suggestion that these seeds might be PBHs forming around $1$~s after the big bang has therefore become popular. However, the seeds could equally well be PCBHs,  left over from the galaxies formed in the previous cosmic cycle. The galaxies themseves would not survive the bounce but the central SMBHs could do so.  
In this case, the same black holes could provide galactic seeds in successive cosmic cycles. 
To be more precise, a PBBBH of mass $m$ provides an initial seed fluctuation for objects of mass $M$
of amplitude $\delta \sim m/M$.
\if
and the Poisson effect ($\delta \sim(f m/M)^{1/2}$) where $f$ is fraction of mass in PBHs \cite{carr-silk2}.
$f \sim 1$ if PBHs provide dark matter but  also consider scenarios with$ f \ll 1$. 
If PBBBHs provide the dark matter ($f \sim 1$), then the Poisson effect is bigger. If $f \ll 1$, the seed effect dominates for $M < m/f$ and locally even for high M, it produces a characteristic density profile
\fi  
This fluctuation grows as $z ^{-1}$ after matter-radiation equality ($z \sim 10^4$), so the mass binding at redshift $z_B$ is $M \sim 10^4 m z_B^{-1} $.
To make a galaxy ($M \sim 10^{12}M_{\odot}$, $z_B \sim 10$), we require $m \sim 10^9M_{\odot}$ or somewhat less if the black hole can  grow through accretion.
 \if
for  seed effect or $m \sim 10^6f^{-1}M_{\odot}$ for Poisson effect. 
To make a Lyman-$\alpha$ cloud ($M \sim 10^{10}M_{\odot}, z_B \sim 10$), we require $m \sim 10^7M_{\odot}$ for seed effect or $m \sim 10^4M_{\odot} /f$ for Poisson effect. Assume SMBHs in AGN are PBHs. 
Both the seed and Poisson effects bind mass $M \propto m$, so 
\fi
This naturally explains why the ratio of bulge mass to SMBH mass is around $10^{3}$, since this is the growth factor of fluctuations between matter-radiation equality and the redshift when SMBHs are in place. In this case, the galaxy mass function in the present cycle just replicates the one in the previous cycle and naturally has the Schechter form.
In this scenario,  the SMBHs serve as a form of ``DNA'' for the transmission of cosmic structure in successive  cycles. This complements  the idea that scale-invariant fluctuations generated in a cosmological collapse phase could generate scale-invariant curvature fluctuations in the next expansion phase \cite{cai}.\\

\item {\it Can the black hole filling factor reach unity?}
The definition of  a black hole in an expanding cosmological background is problematic. One cannot use the concept of an event horizon   (since there may be no spatial infinity) and the issue is further complicated in a bouncing model (since there is no cosmological singularity).  
One must therefore use the concept of the  {black hole} apparent horizon, defined as the outermost Marginally Outer Trapped Surface (MOTS) \cite{MOTS}. 
\if
This approach is now used routinely in numerical work and we will also adopt it here. It has even been argued that Hawking radiation is associated with the apparent horizon \cite{cai}. For a more detailed discussion of these issues, see Ref.~\cite{faraoni}. 
A matter of particular interest for models of this type concerns what happens to the number density of black holes as the minimum of expansion is approached. One could speculate about the maximum {fraction} of space that can be filled by PCBHs or {about} the number of BCBHs that form during the collapsing stage. These quantities would depend on the density of the matter at the bounce and there are then various interesting constraints that can be imposed on the fraction of the universe's mass density that resides in black holes as a function of mass \cite{cc}. If the black holes are randomly distributed, one would expect a process of hierarchical merging to occur, in which progressively larger apparent horizons form around groups of holes, so that the characteristic hole size steadily increases. However, if the holes were distributed uniformly (eg. on a lattice), then pre-crunch hierarchical merging would be eliminated and they might maintain their individuality. In this case, one could investigate the maximum number of black holes that could fit through the bounce.
What happens when the Schwarzschild radius of the black hole becomes
comparable to the size of the cell,  i.e. when  the filling factor $f$
tends to unity?  
In considering the possibility of PCBHs, the crucial question is whether  the volume filling factor of the black holes $f$ exceeds $1$ at the time of the bounce. 
\fi
Given this definition of the size of a black hole, one can ask what happens as the volume filling factor of PBBBHs approaches $1$. 
 If they are randomly distributed, one would expect a process of hierarchical merging to occur, in which progressively larger horizons form around groups of holes, so that the characteristic hole size steadily increases. However, if they have a precisely uniform lattice distribution, one could envisage the whole universe suddenly forming a single black hole at some epoch, in the sense that the individual horizons disappear.
\if
 thereafter  the background would appear nearly homogenous to every observer but with a ``sprinkling'' of singularities. Since all these singularities are connected as part of the Big Crunch singularity and lie in the future as part of $i^+$ (rather than being naked), 
and therefore are not naked singularities. So in what are they a sprinkling of singularities? Although
\fi 
The interpretation of this situation is unclear. 
C$^2$ addressed this question on the assumption that the volume filling factor  $f$ is of order $(R_S/L)^3$, where $R_S$ is the Schwarzschild radius of the black holes
and $L$ is their separation at the bounce. C$^3$ calculated $f$ more precisely in the context of their exact cosmological models and found that there are indeed solutions in which the universe bounces before $f$ reaches $1$. 
\if
However, they found that $f$ never reaches $1$ in the 3D situation because $M$ falls to $0$ and then becomes negative as $f$ inreases above $0.8$. These solutions are presumably unphysical. On the other hand, 
$a$ has a maximum rather than a miniumum) for $f < 0.4$.  
In the $n$-dimensional case  there do exist bouncing solutions with positive $M$ in which $f$ can reach $1$ for $n = 4, 5, 8, 9$ but not $n = 6, 7$.
there are bouncing solutions in which $f$ exceeds $1$, although it was unclear whether the models were mathematically self-consistent in this situation.]

The first is that the singularities appear as Planck-mass objects which  might might even provide the dark matter.  Such Planck mass relics are usually regarded as the end-state of Hawking evaporation and provide an intriguing dark matter candidate, since they are midway between MACHOs and WIMPs. As regards their mass, they are much smaller than MACHOs but much larger WIMPs. As regards their size, they are the smallest conceivable object in the Universe ($R_P \sim 10^{-33}$cm), which means that they are virtually undetectable except through ther gravitational effects.
\fi
Note that Penrose \cite{penrose} would discount  all the black holes in the universe merging into a single black hole of mass $M_U$ during the big crunch on the grounds that the entropy would then be $S \sim M_U^2$ in Planck units, which is larger than the entropy in the radiation by a factor of $10^{-27} M_U > 10^{33}$. Indeed, he uses this to argue that there cannot have been a previous collapse phase. However, there might be ways to avoid this conclusion. One possibility is that the filling factor never reaches $1$ because the size of the black hole horizon
 becomes less than the usual Schwarzschild value $R_S$ as $f$ increases. After all, the horizon size is expected to be modified in a cosmological background.
However, we have no rigorous proof that $f$ must always remain below $1$.\\

\item {\it Thermodynamics of cyclic model.}
Tolman \cite{tolman}  argued from thermodynamic considerations that the amplitude and period of a closed cyclic universe must grow at each bounce as a result of the steady increase of entropy. On this basis, it has been argued that if we live in an oscillatory universe  we can be at most 100 cycles away from the first cycle that lasted long enough to produce stars \cite{dicus}. 
If we live in a cyclic universe with more than one bounce, we may ask whether any quantity changes systematically from one cycle to the next  due to the persistence of black holes. For example, one would expect the number of black holes 
to increase with each cycle, corresponding to an increase in the gravitational entropy.
In this case, once  the filling factor increases to unity, 
the cycles may terminate.\\

\if
But this process might cease once the filling factor reaches unity. Also of interest is what happens in the past. Do the cycles continue to the infinite past? If the filling factor decreases for some period, maybe the cycles begin when the filling factor is unity. \\
ikkema and Israel \cite{sik} resolve this paradox by considering the interior properties of a rotating Kerr black hole. the inner horizon has an infinite blue shift for infalling radiation. So the tail of gravitational waves generated by the collapse will grow to Planck densities before quantum damping becomes important, implying that the gravitational mass inflates to $m_{core} \sim m^3$ in Planck units.  For $M \sim 5 M_{\odot}$, this is $10^{55}M_U$, although is not measured by an outside observer. Since angular mementum does ot increase, it should be well described by the Schwarzschild solution. This deceases the time of collapse from the Schwarzschild time to the Planck time. The Universe undergoes a phase transition to Planckian density. Time deflation brings all cores to a common zero of time to within a Planck time. The totla entropy remans sub-Plankcain becasuse the Bekenstein-Hawking formula no longer applies. During the Planckian bounce, the Wheeler-de Witt equation implies that one has regeneration of order. The re-emerging universe is very homogenous on scales much larger than the horizon, so this resolves the horizon problem. If the Universe is closed, its origin can be traced back to a first ``baby'' universe large enough to produce black holes whose evaporation time allowed them to survive unitl the crunch. This period may be crucial to understanding the origin of the dimensionality of the universe and the value of the cosmological constant.   
\fi

\item {\it Higher-dimensional models.}
It is possible that the universe becomes higher-dimensional 
as it approaches the big crunch. This is the time-reverse of the dimensional reduction which may occur in the expansion phase. 
In particular, in brane cosmology the universe is regarded as a $4$-dimensional brane in a $5$-dimensional bulk \cite{Maartens}. However, at times sufficiently early that the horizon scale is less than the brane thickness or compactification scale, the universe becomes effectively $5$-dimensional. Indeed, these models can be described by the 5-dimensional Schwarzschild - de Sitter solution, in which the universe effectively emerges from a 5-dimensional black hole \cite{muk}. So maybe the filling factor reaching unity corresponds to the network of 4-dimensional black holes merging into a single 5-dimensional one. 
There could even be a hierarchy of compactification scales, so that the dimensionality of the universe progressively increases as one goes back to earlier times. The scenario is radically changed in this case and this was the motivation for considering higher-dimensional models in Ref.~\cite{ccc}. 
\if
As we now discuss, there is a sense in brane cosmology in whihc necessarily necessarily becomes a $5$-dimensonal black hole.
Finally, we considered the bounce and merger conditions on persistent black holes through a cosmological bounce in higher-dimensional spaces, which may be of relevance in cosmological scenarios in which the number of spatial dimensions in the Universe increase at sufficiently high densities. 
The lower bound on the values of the model parameter $\kappa_n$ required for a bounce seems to increase with the number of dimensions. However, it appears that that there are no bouncing cosmologies containing non-merging black holes of the type discussed in this paper if there are more than five spatial dimensions (i.e. if $n>5$).  We also found that the upper bound on the number density of black holes that exists in $n=3$, {coming} from the requirement that the mass of each black hole {be} positive, is removed when $n=4$ or $5$.
\fi

\end{itemize}

\if
In this essay we have considered bouncing cosmological models at the moment of maximum compression. Our models contain a lattice of black holes in a Universe whose energy density is dominated by a scalar field. We then obtained exact solutions for time-evolving models in which multiple distinct black holes persist through the bounce.
In this essay we have explored some of the cosmological consequences of the possibility that black holes can persist in a universe that collapses to a big crunch and then bounces into a new expansion phase. We use a scalar field to model the matter content of such a universe {near the time} of the bounce and demonstrate 
that there exist models in which multiple distinct black holes can persist through a bounce. This allow fconcrete computations of quantities such as the black hole filling factor and infer conditions for the black holes to remain distinct {(i.e. avoid merging) and hence} persist into the new expansion phase.. We consider solutions in flat cosmologies, as well as in higher-dimensional spaces (with up to nine spatial dimensions).  .
\fi

There are several areas in which the current analysis could be extended. This includes the study of models in which the black holes persist {without the bounce being time-symmetic, as might be more realistic if the entropy increases at a bounce, and  more {detailed} higher-dimensional models. 
\if
Beyond these more mathematical questions, there are also a number of potentially very interesting physical consequences of the models presented here.
For example, if there is more than one bounce then it is possible that there are more black holes in each subsequent cycle (i.e. both the newly created ones and the persisting {ones from previous cycles). In this case}, the filling factor of black holes increases with each cycle, so that the bounces may at some point terminate.
\fi 
But it is clear that the persistence of black holes through a cosmological bounce could be relevant to many cosmological conundra.

\end{document}